\documentclass{article}
\usepackage{spconf,amsmath,graphicx}
\usepackage{amsfonts}
\usepackage{hyperref}
\usepackage{color}
\usepackage{setspace}
\usepackage{tikz}

\newcommand\copyrighttext{%
	\footnotesize \textcopyright 2020 IEEE. Personal use of this material is permitted.
	Permission from IEEE must be obtained for all other uses, in any current or future 
	media, including reprinting/republishing this material for advertising or promotional 
	purposes, creating new collective works, for resale or redistribution to servers or 
	lists, or reuse of any copyrighted component of this work in other works. 
	DOI: \href{https://doi.org/10.1109/ICIP42928.2021.9506763}{10.1109/ICIP42928.2021.9506763}}
\newcommand\copyrightnoticeOwn{%
	\begin{tikzpicture}[remember picture,overlay]
		\node[anchor=north,yshift=-10pt] at (current page.north) {\fbox{\parbox{\dimexpr\textwidth-\fboxsep-\fboxrule\relax}{\copyrighttext}}};
	\end{tikzpicture}%
\vspace{-8mm}
}

\title{Analysis of Neural Image Compression Networks for Machine-to-Machine Communication}

\name{Kristian Fischer, Christian Forsch, Christian Herglotz, and Andr\'e Kaup\thanks{The authors gratefully acknowledge that this work has been supported by the Deutsche Forschungsgemeinschaft (DFG) under contract number KA~926/10-1.}}
\address{Multimedia Communications and Signal Processing\\
	Friedrich-Alexander-Universit\"at Erlangen-N\"urnberg (FAU)\\
	Cauerstr. 7, 91058 Erlangen, Germany\\
	\{Kristian.Fischer, Christian.Forsch, Christian.Herglotz, Andre.Kaup\}@fau.de}
\begin{document}
%
\maketitle
\copyrightnoticeOwn
\begin{abstract}

Video and image coding for machines~(VCM) is an emerging field that aims to develop compression methods resulting in optimal bitstreams when the decoded frames are analyzed by a neural network. Several approaches already exist improving classic hybrid codecs for this task. However, neural compression networks~(NCNs) have made an enormous progress in coding images over the last years. Thus, it is reasonable to consider such NCNs, when the information sink at the decoder side is a neural network as well. Therefore, we build-up an evaluation framework analyzing the performance of four state-of-the-art NCNs, when a Mask R-CNN is segmenting objects from the decoded image. The compression performance is measured by the weighted average precision for the Cityscapes dataset. Based on that analysis, we find that networks with leaky ReLU as non-linearity and training with SSIM as distortion criteria results in the highest coding gains for the VCM task. Furthermore, it is shown that the GAN-based NCN architecture achieves the best coding performance and even out-performs the recently standardized Versatile Video Coding~(VVC) for the given scenario.

\end{abstract}
\section{Introduction}
\label{sec:intro}

Throughout the recent decades, image and video compression has been dominated by classic hybrid coding methods like Joint Picture Experts Group (JPEG)~\cite{wallace1992_jpeg}, High Efficiency Video Coding~(HEVC)~\cite{sullivan2012_HEVC}, and Versatile Video Coding~(VVC)~\cite{chen2020vtm10}. But with the rise of neural networks, multiple methods were proposed to train neural image compression networks~(NCNs) end-to-end by balancing the contrary goals of a small bitstream and best possible image quality~\cite{balle2018_b2018, balle2018_bmshj2018, minnen2018_mbt2018, mentzer2020hific}. Thereby, all those networks already provide a superior rate-distortion performance than JPEG coding.

Another consequence of the tremendous advances in the field of neural networks is that more and more applications in everyday life, which perform tasks from the field of computer vision, are based on neural networks. Such networks are applied, e.g., for video surveillance, industrial processes, and autonomous driving. In most real-world applications, the multimedia data has first to be transmitted or stored from the capturing device before being analyzed by the neural network. This requires a suitable compression scheme, which is usually optimized for providing the best possible quality for the human visual system. But, as shown in~\cite{fischer2020_ICIP}, this does not always have to result in a high coding performance, when the decoded frame is analyzed by a neural network instead.
Optimizing codecs such that the decoded frame can optimally be analyzed by a neural network is attributed to the field of video coding for machines~(VCM), which is targeted by the MPEG ad-hoc group~\cite{zhang2019} founded in 2019. Besides, several other work was proposed designing or optimizing coding chains with classic hybrid codecs for such machine-to-machine~(M2M) communication~\cite{bagdanov2011, dodge2016, galteri2018,fischer2020_FRDO}. 

\begin{figure}
	\centering
	\includegraphics[width=0.45\textwidth]{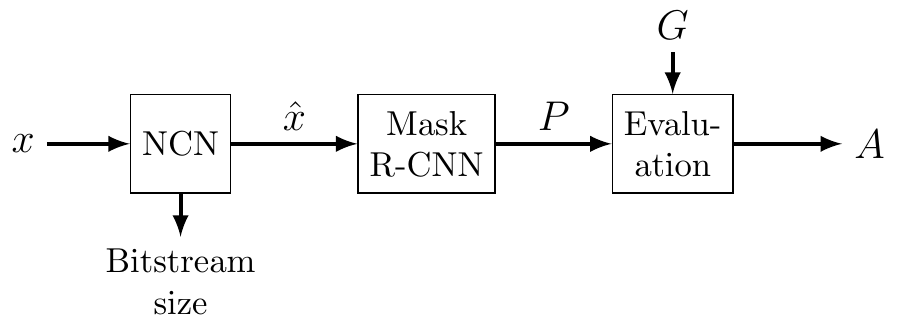}
	\caption{Investigated framework coding input image $x$ for neural networks. $A$ denotes the accuracy of Mask R-CNN segmenting objects.}
	\label{fig: investigated framework}
	\vspace{-4mm}
\end{figure}

Derived from the two before-mentioned developments, this paper deploys NCNs for the VCM task for the first time, investigating which architectures and parametrizations of NCNs are best suited for the VCM task. This provides valuable information to reach the ultimate objective of training such networks end-to-end for M2M communication. 

For the investigations, we build up an M2M scenario as shown in Fig.~\ref{fig: investigated framework}, where the decoded frame $\hat{x}$ is analyzed by the state-of-the-art instance segmentation network Mask R-CNN~\cite{he2017}. Its accuracy $A$ is measured by comparing the detections $P$ against the ground truth $G$ over the required bitrate to obtain the coding efficiency of the investigated NCNs. Thereby, several architectures and methods of neural compression networks are tested including different non-linearities~\cite{balle2018_b2018}, different distortion metrics during the training process~\cite{balle2018_bmshj2018, minnen2018_mbt2018}, and a Generative Adversarial Network~(GAN)~\cite{goodfellow_2014_gan} structure~\cite{mentzer2020hific}. Besides, their performance is compared in relation to the commonly used perceptual distortion metrics PSNR, Structural Similarity index (SSIM)~\cite{wang2004}, and Video Multi-method Assessment Fusion~(VMAF)~\cite{li2016_vmaf_short}. In the final experiment, we compare the investigated neural compression networks against JPEG and the state-of-the-art video compression methods HEVC and VVC applied in all-intra configuration. 

\begin{figure}
	\centering
	\includegraphics[width=0.49\textwidth]{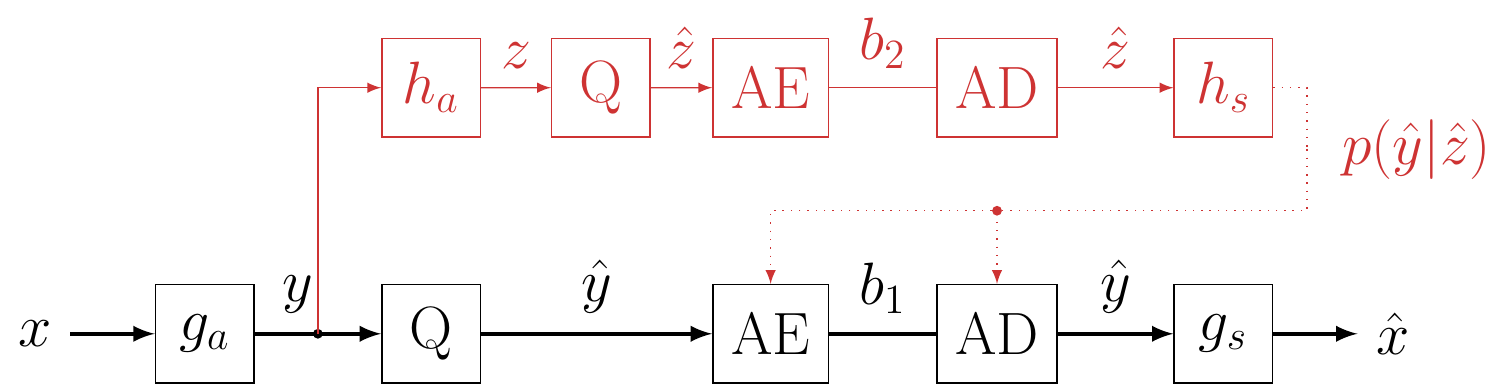}
	\caption{\textit{b2018} architecture in black; Additional hyperprior coding structure for \textit{bmshj2018} is depicted in red. AE and AD denote lossless arithmetic en- and decoding, respectively.}
	\label{fig: bmshj2018 build-up}
	\vspace{-3mm}
\end{figure}

\section{Investigated Image Compression Networks}
\label{sec:investigated image compression networks}

\subsection{Basic Neural Compression Network -- \textit{b2018}}
In hybrid image or video codecs, transform coding is deployed to reduce statistical dependencies by transforming the residual image into a frequency domain. Subsequently, the coefficients are quantized and encoded by an entropy encoder to reduce the bitrate. As transformation, the linear Discrete Cosine Transform (DCT) is commonly selected and non-linear methods such as prediction are added to improve the performance for non-linear signals. 

Contrary, for neural compression networks as depicted in Fig.~\ref{fig: bmshj2018 build-up} and proposed in~\cite{balle2017endtoend} and~\cite{balle2018_b2018}, the transform is directly implemented as an analysis neural network $y = g_a(x, \phi)$ generating the latent space $y$ from the input image $x$ with the learned network weights $\phi$. Subsequently, the latent space $y$ is quantized into $\hat{y}$. By encoding $\hat{y}$ losslessly and transmitting it to the decoder side, the decoded output image $\hat{x}$ is obtained by applying an inverse synthesis transformation $\hat{x} = g_s(\hat{y}, \theta)$ parametrized by $\theta$. The weights $\phi$ and $\theta$ are jointly trained by minimizing the loss function 
\begin{equation}
\label{eq:loss function}
	\mathcal{L}(\phi,\theta,\psi)= \mathbb{E}_{x\sim p_x} [-\log_2 p_{\hat{y}}(\hat{y},\psi) + \lambda \cdot d(x, \hat{x})].
\end{equation}
Thereby, the first summand holds the entropy $H$ of $\hat{y}$ with the estimated entropy model $p_{\hat{y}}$ and its parametrization $\psi$. The lower the entropy, the less bits $b$ will be required to transmit $\hat{y}$ to the decoder side. The second summand represents the distortion between the original image $x$ and its reconstructed version $\hat{x}$ measured by an arbitrary distortion function $d$. Typically, Mean Squared Error~(MSE) is chosen as $d$. Similar to hybrid coding methods, $\lambda$ steers the relaxation between low bitrate and high quality towards either direction.

Both transform networks, $g_a$ and $g_s$, build an hourglass-shaped auto encoder structure to derive a latent space $y$ with a lower dimensionality than $x$ in order to achieve a more compact representation that can efficiently be transmitted to the decoder. They consist of convolution layers adapting the spatial resolution with a down- or upscaling stride.
Subsequently, a Generalized Divisive Normalization~(GDN)~\cite{balle2016density} non-linearity is applied, which is inspired from visual systems occurring in nature and increases statistical independence by normalizing inside a layer. Another non-linearity alternative is the leaky Rectified Linear Unit (LReLU), which is often used in classification and detection networks.

\vspace{-1mm}
\subsection{NCN with Additional Hyperprior -- \textit{bmshj2018} and \textit{mbt2018}}
One major drawback of the architecture in \textit{b2018} is that the latent space $\hat{y}$ still holds spatial dependencies. Thus, a hyperprior is added to the \textit{b2018} architecture in~\cite{balle2018_bmshj2018} and shown in Fig.~\ref{fig: bmshj2018 build-up}, which includes a second auto encoder consisting of the analysis and synthesis networks $h_a$ and $h_s$, respectively. This auto encoder obtains the statistical dependencies $p(\hat{y}| \hat{z})$ from a second latent space $z$. Thus, each element $\hat{y}_i$ can be modeled by a Gaussian distribution with zero mean and the standard deviation being derived from this additional latent space $\hat{z}$. With this \textit{bmshj2018} model, the coding performance is significantly increased over $b2018$. Additionally, the \textit{bmshj2018} model is also proposed to be trained with the multi-scale SSIM metric as distortion function $d(x,\hat{x})$.

\begin{figure}
	\centering
	\includegraphics[width=0.49\textwidth]{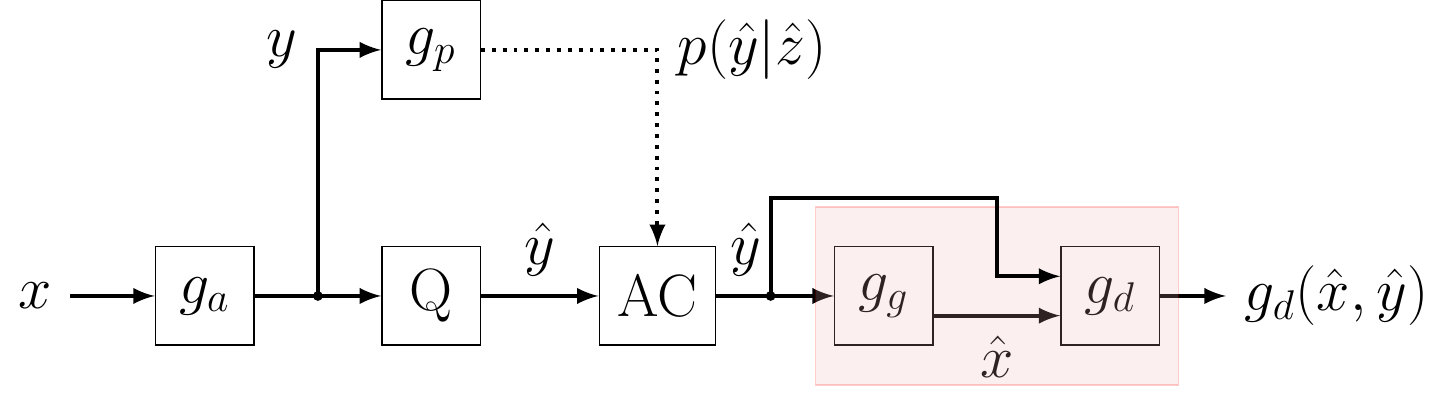}
	\vspace{-6mm}
	\caption{\textit{HiFiC} architecture; Arithmetic coding~(AC) comprises AE and AD for simplicity. The red highlighted area represents GAN structure at the decoder side.}
	\label{fig: hific build-up}
	\vspace{-3mm}
\end{figure}

The successor of \textit{bmshj2018}, \textit{mbt2018}~\cite{minnen2018_mbt2018}, employs non-zero-mean Gaussian distributions to code $\hat{x}$. Additionally, an autoregressive context model is added to \textit{bmshj2018} to further improve the entropy coding step.

\subsection{GAN-based NCN -- \textit{HiFiC}}
The last considered model \textit{High Fidelity Compression (HiFiC)} \cite{mentzer2020hific} is a GAN-based expansion of \textit{bmshj2018}. Its basic structure is depicted in Fig.~\ref{fig: hific build-up}. Similar to \textit{bmshj2018}, the encoding network $g_a$ generates a quantized latent space $\hat{y}$, which is entropy coded with the help of an additional hyperprior $\hat{z}$ derived from $g_p$. The decoder is designed as a GAN being conditioned on $\hat{y}$. There, the generator network $g_g$ is supposed to fool the discriminator network $g_d$ by creating output images $\hat{x}$ derived from $\hat{y}$ that $g_d$ falsely classifies as real world data, which results in superior subjective quality than \textit{bmshj2018} or \textit{mbt2018}.
Besides, \textit{HiFiC} is additionally trained with a distortion metric measured in a feature space of a neural network~\cite{zhang2018_LPIPS_metric}, which was also shown to be beneficial for VCM coding with VVC and Mask R-CNN in~\cite{fischer2020_FRDO} by a similar metric.

\section{Analytical Methods}
\subsection{Dataset}
In order to evaluate a framework as shown in Fig.~\ref{fig: investigated framework}, the 500 uncompressed images with a size of $1024\times 2048$ pixels from Cityscapes~\cite{cordts2016} validation set are encoded with the different neural compression networks. These images are captured from a car's windshield observing different road scenes. For each image, a pixel-wise annotation of eight different classes of road users is provided. With that, the Average Precision~(AP) is calculated as proposed for Cityscapes~\cite{cordts2017cityscapesscripts} over the whole dataset and for each class in order to measure the Mask R-CNN accuracy. Ultimately, the AP values are weighted~(wAP) according to the number of instances for each class as proposed in~\cite{fischer2020_ICIP} and~\cite{fischer2020_FRDO}.

\subsection{Employed Implementations}

To compress the Cityscapes images at full resolution, we utilize the pre-trained NCN models provided by the original authors in~\cite{git_tensorflowCompression} without further re-training.
For \textit{bmshj2018} and \textit{mbt2018}, eight models exist covering different areas of the rate-distortion relaxation, whereas for \textit{b2018} and \textit{HiFiC} only four and three models are supplied, respectively.

As state-of-the-art hybrid intra video coding reference, the HEVC test model~(HM 16.20)~\cite{hm_software} and VVC test model~(VTM 10.0)~\cite{chen2020vtm10} are selected. Before applying these two codecs, the Cityscapes images provided as PNGs are first converted into YUV format with 4:2:0 downscaling and vice-versa before applying Mask R-CNN. In order to fit to the bitrate ranges provided with the NCN models, Quantization Parameter (QP) values of 12 to 42 in steps of 5 are chosen. Lastly, JPEG compression is investigated using the OpenCV library~\cite{opencv_library} with quality levels from 10 to 90 in steps of 10.

To detect the road users from the compressed images, the Detectron2~\cite{wu2019detectron2} framework is deployed. It provides a Mask R-CNN model with a ResNet-50~\cite{he2016resnet} backbone that has already been trained on the Cityscapes training images.

\vspace{-2mm}
\subsection{Quality Metrics}
In order to measure the performance of the different codecs for the human visual system, the quality metrics PSNR, SSIM, and VMAF are obtained. The wAP of Mask R-CNN being applied to the compressed images is taken to measure the performance for the M2M scenario. In order to quantify the resulting rate-distortion curves, the Bj\o ntegaard delta rate~(BDR)~\cite{bjontegaard2001_new} is calculated, which measures the bitrate savings for an identical quality. In addition to common BDR using PSNR, SSIM, and VMAF as quality metric, PSNR is also substituted with wAP to measure the VCM coding performance as it is recommended by MPEG VCM group~\cite{liu2020_VCM_CTC}.

\vspace{-2mm}
\section{Evaluation Results}
\vspace{-2mm}
\subsection{Choice of Non-Linearity}
\label{subsec: choice of non-linearity}
The first experiment conducts a comparison between a \textit{b2018} model build with GDN non-linearities that are optimized for compressing natural content for the human visual system and a model build with LReLUs. The BDR values for the different quality metrics are provided in Table~\ref{tab: BDR b2018}. Choosing LReLU over GDN as non-linearity requires more bits to achieve the same PSNR. Contrary, the coding performance when coding for Mask R-CNN is significantly improved by selecting the LReLU model, which saves 13.2\,\% bitrate for the same wAP. Similar ReLU activations are also used in the ResNet backbone of Mask R-CNN, which is one possible explanation, why the LReLU-based \textit{b2018} model outperforms the GDN-based model for M2M communication.

\begin{table}[]
	\caption{BDR in \% with respect to the listed quality metric using \textit{b2018} with GDN as anchor for four quality levels.}
	\label{tab: BDR b2018}
	\centering
	\vspace{1mm}
	\begin{tabular}{l|llll}
		\hline
		                     & PSNR & VMAF & SSIM & wAP   \\ \hline
		\textit{b2018-LReLU} & 1.9  & -2.4 & 2.0  & -13.2 \\ \hline
	\end{tabular}
\vspace{-3mm}
\end{table}

\subsection{Influence of Training Distortion Metric}
\label{subsec: influence of used distortion-metric}

\begin{table}[]
	
	\caption{BDR in \% with respect to the listed quality metric using the corresponding codec trained with MSE as anchor for eight quality levels.}
	\label{tab: BDR SSIM}
	\centering
	\vspace{1mm}
	\begin{tabular}{l|llll}
		\hline
		& PSNR & VMAF & SSIM & wAP   \\ \hline
		\textit{bmshj2018-SSIM} & 32.2  & 18.5  & -35.2  & -6.3 \\
		\textit{mbt2018-SSIM} & 46.7  & 36.6  & -31.1  & -1.0 \\ \hline
	\end{tabular}
\vspace{-3mm}
\end{table}

Another important influence on the performance of neural compression networks is the selected distortion metric throughout the training process. Here, the performances of the two compression networks \textit{bmshj2018} and \textit{mbt2018} are compared depending on whether they were trained with MSE or SSIM. The BDR results are listed in Table~\ref{tab: BDR SSIM}. Naturally, training the models on SSIM performs worse when measuring the output quality with PSNR as well as for VMAF, but immensely increases the coding performance with respect to SSIM. Having the goal to achieve a high detection accuracy measured by wAP, the model should be trained with SSIM as distortion metric as well, saving up to 6.3\,\% and 1.0\,\% of bitrate, respectively. This can have multiple explanations. First, the model trained with SSIM focuses, as well as the evaluation network Mask R-CNN, on the structural information~\cite{wang2004} of the content. Second, the \textit{bmshj2018} authors stated that their model trained on SSIM focuses on regions of low contrast by omitting information in high contrast areas. This accommodates the Mask R-CNN, which is struggling to segment objects that do not differ much from the background because they are for example located in the shadow of a building, which can occur throughout the Cityscapes dataset, and which gets amplified when adding quantization to the image.


\subsection{Comparison of Neural Compression Networks against State-of-the-Art Compression Methods}
\label{subsec: Comparison of Neural Network Compression Algorithms Against State-of-the-Art Compression Methods}

\def\plotWidth{0.4}
\begin{figure}[t]
	\centering
	\includegraphics[width=\plotWidth\textwidth]{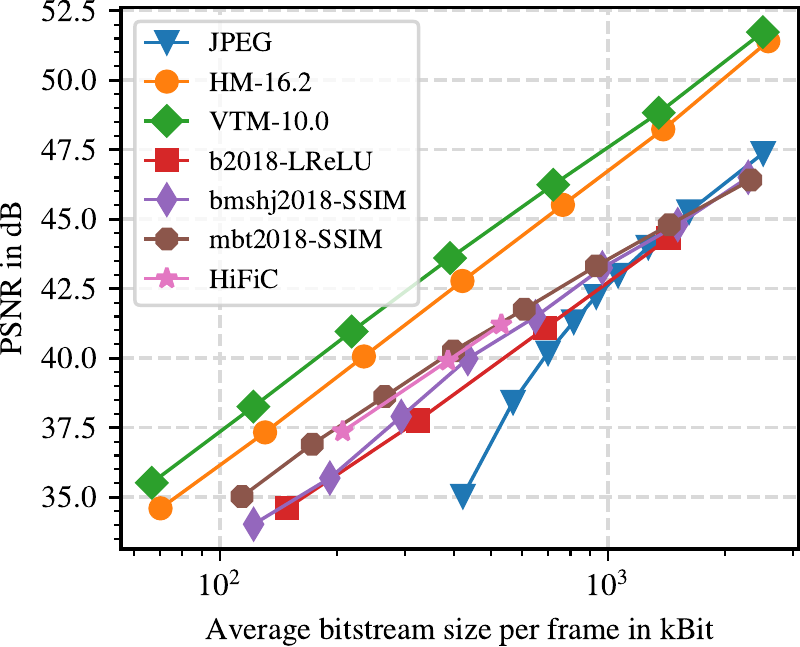}
	\vspace{-3mm}
	\caption{PSNR over required bits for investigated codecs and the parametrization resulting in the best performance for VCM task and for the 500 Cityscapes validation images.}
	\label{fig:PSNR over rate}
	\vspace{-4mm}
\end{figure}

\begin{figure}[t]
	\centering
	\includegraphics[width=\plotWidth\textwidth]{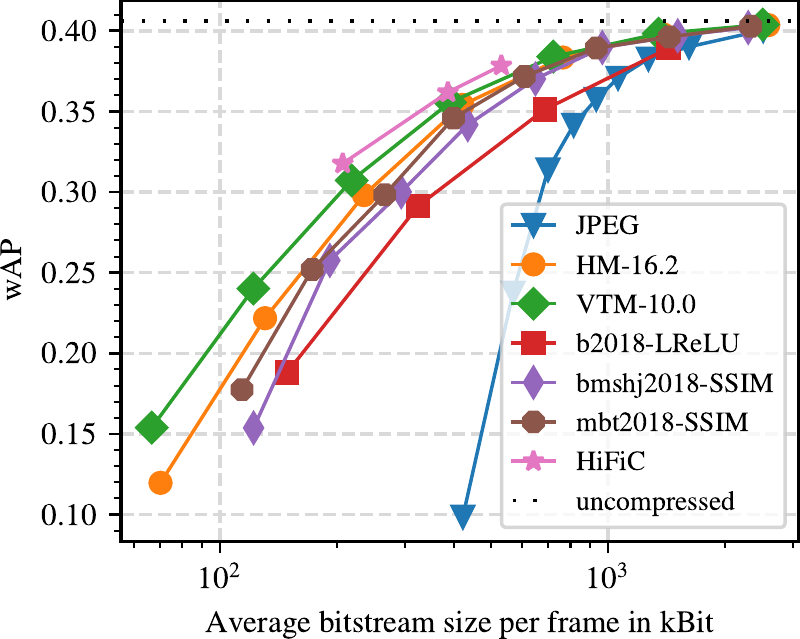}
	\vspace{-3mm}
	\caption{wAP over required bits for investigated codecs and the parametrization resulting in the best performance for VCM task and for the 500 Cityscapes validation images. The black dotted line represents the accuracy when applying Mask R-CNN to the uncompressed Cityscapes images.}
	\label{fig:wAP over rate}
	\vspace{-2mm}
\end{figure}

In the final analysis, all chosen models from Section~\ref{sec:investigated image compression networks} with their best found parametrization are compared against the classic codecs JPEG, HEVC, and VVC. Figures~\ref{fig:PSNR over rate} and~\ref{fig:wAP over rate} provide the rate-PSNR and rate-wAP curves, respectively. Table~\ref{tab: BDR all codecs} lists the BDR of all codecs with the \textit{b2018-LReLU} model as anchor. Among the NCNs, \textit{mbt2018} with the corresponding training distortion metric performs best for PSNR, VMAF, and SSIM. However, the classic video codecs HEVC and VVC still achieve higher BDR savings.

Regarding the investigated VCM use case, \textit{HiFiC} outperforms all other codecs, even performing better than the upcoming video coding standard VVC. The reason for this seems to be caused in the GAN-based structure of \textit{HiFiC} and the neural-network-based distortion metric. During training, the network is pushed towards producing compressed images $\hat{x}$ resulting in a high activation of the discriminator network $g_d$. That can be compared to the investigated VCM inference case providing images $\hat{x}$ that result in the best possible detection and segmentation accuracy of Mask R-CNN.

\begin{table}[]
	\caption{BDR in \% with respect to the listed quality metric using \textit{b2018} with LReLU and trained on MSE as anchor. Highest bitrate savings for each quality metric is set in bold.}
	\label{tab: BDR all codecs}
	\centering
	\vspace{1mm}
	\begin{tabular}{l|rrrr}
		\hline
		                          & PSNR  & VMAF  & SSIM  & wAP   \\ \hline
		JPEG             		  & 38.0  & -6.5  & 41.6  & 96.6  \\
		HM-16.20                   & -56.5 & -61.3 & -49.7 & -33.5 \\ 
		VTM-10.0                  & \textbf{-66.8} & \textbf{-71.3} & \textbf{-60.5} & -43.2 \\
		\textit{bmshj2018-MSE}    & -32.0 & -44.5 & -23.1 & -15.5 \\
		\textit{bmshj2018-SSIM} & -11.9 & -33.5 & -53.5 & -21.3 \\
		\textit{mbt2018-MSE}      & -49.2 & -55.3 & -41.6 & -26.9 \\
		\textit{mbt2018-SSIM}   & -28.6 & -39.8 & -62.4 & -27.9 \\
		\textit{HiFiC}            & -27.7 & -49.4 & -51.4 & \textbf{-52.8} \\ \hline
	\end{tabular}
\end{table}

\section{Conclusions}
This paper analyzed several neural compression networks according to their performance, when Mask R-CNN is applied to analyze the compressed images. The experiments first revealed that the \textit{b2018} model with LReLU as activation function achieved a superior wAP-rate performance than the GDN-based model. Besides, training models with SSIM was shown to result in bitrate savings compared to standard training with MSE as distortion metric, when coding for Mask R-CNN. Moreover, the GAN-based network \textit{HiFiC} outperformed all other NCNs and the state-of-the-art codecs for the given scenario. Additional experiments in future might find whether this is mostly caused by the GAN structure or the feature-based distortion metric applied in training. Derived from these promising results, future work will now aim for superior NCN coding performance for machines. This could be achieved by improving the training process with enhanced error metrics representing the behavior of image analysis networks and end-to-end training with Mask R-CNN.


\newpage
\bibliographystyle{IEEEbib}
\setstretch{1}
\small
\bibliography{/home/fischer/Paper/jabref_literature_research_ms2.bib,/home/fischer/Paper/literature_M2M_communication.bib,/home/fischer/Paper/jabref_used_software.bib}

\end{document}